\documentclass[aps,showpacs,twocolumn]{revtex4}
\usepackage{amsfonts}
\usepackage{amsmath}
\usepackage{amssymb}
\usepackage{graphicx}

\begin{document}

\title{Asymmetric Field Photovoltaic Effect of Neutral Atoms}
\author{Wenxi Lai$^{1}$} \author{Jinyan Niu$^{2}$} \author{Yu-Quan Ma$^{1}$} \author{W. M. Liu$^{3,4}$}\email{wliu@iphy.ac.cn}
\affiliation{1 School of Applied Science, Beijing Information Science and Technology University, Beijing 100192, China}
\affiliation{2 School of Science, Inner Mongolia University of Science and Technology, Baotou, 014010, China}
\affiliation{3 Beijing National Laboratory for Condensed Matter Physics, Institute of Physics, Chinese Academy of Sciences, Beijing 100190, China}
\affiliation{4 School of Physical Sciences, University of Chinese Academy of Sciences, Beijing 100190, China}

\begin{abstract}
Photovoltaic effect of neutral atoms using inhomogeneous light in double-trap opened system is studied theoretically. Using asymmetric external driving field to replacing original asymmetric chemical potential of atoms, we create polarization of atom population in the double-trap system. The polarization of atom number distribution induces net current of atoms and works as collected carriers in the cell. The cell can work even under partially coherent light. The whole configuration is described by quantum master equation considering weak tunneling between the system and its reservoirs at finite temperature. The model of neutral atoms could be extended to more general quantum particles in principle.
\end{abstract}


\maketitle

\begin{center}
\textbf{I. INTRODUCTION}
\end{center}

Research on new devices for future atomtronic circuit have attracted great attentions recently, such as atomic clocks~\cite{Katori,Ludlow,Schioppo}, atom interferometry~\cite{Cronin,Hamilton,Gebbe}, atom transistors~\cite{Fuechsle,Caliga2}, atom chips~\cite{Folman,Riedel,Bernon}, quantum logic gates~\cite{Sorensen,Calarco,Saffman,Safaei} and atomic batteries~\cite{Seaman,Caliga,Zozulya,wxlai}. They reveal that atoms are well controllable and have substantial degrees of freedom, although they work under critical environments of low temperature at present. Atomic batteries are one kind of these important devices which could be applied to supply power to the others.

Prototype batteries of atoms have been reported both experimentally and theoretically. As far as we know, they include batteries based on chemical potential difference~\cite{Seaman,Caliga}, asymmetric trap~\cite{Zozulya} and artificial gauge fields induced spin-orbit coupling~\cite{wxlai}. In the chemical potential difference based atomic battery, the chemical potential difference has been defined as the effective voltage of the cell~\cite{Seaman}. Such battery can be charged and discharged using a sweeping barrier of radio-frequency field in a double-well structured magnetic chip trap~\cite{Caliga}. In the asymmetric trap configuration, power of the battery comes from the non-equilibrium process between noncondensed thermal atoms and Bose-Einstein condensed atoms with an incoming beam of cold atoms in a highly asymmetric potential~\cite{Zozulya}. In the spin-orbit coupling induced photovoltaic cell, a coherent light inputs energy into cold atoms due to spin-orbit coupling~\cite{wxlai}. Atom population at excited energy level in the double-trap cell can be controlled by the artificial magnetic flux in synthetic dimensional space of atoms.

In this paper, we would demonstrate atomic battery based on asymmetric driving field by trapping atoms in a symmetric potential. Asymmetric field here means two optical fields acting on double traps respectively with different amplitudes, different frequencies or different phases. The phase of light in the present model can be any phase which may be caused by path length of light, initial phase~\cite{Meystre}, complex dipole moment of atoms~\cite{Scully} or artificial magnetic flux induced by spin-orbit coupling~\cite{Gerbier,Wall,Celi,Mancini,Livi}. Therefore, it would be naturally proved in the following that the artificial gauge field induced photovoltaic effect in which phase effect has been considered~\cite{wxlai} is just one particular case of the present model. In fact, current caused by the amplitude and frequency difference is much larger than the current created due to the phase difference. It indicates that amplitude and frequency of light can play more important role than the phase in this kind of photovoltaic cell.

\begin{center}
\textbf{II. THEORETICAL MODEL}
\end{center}

\begin{figure}
\includegraphics[width=8cm]{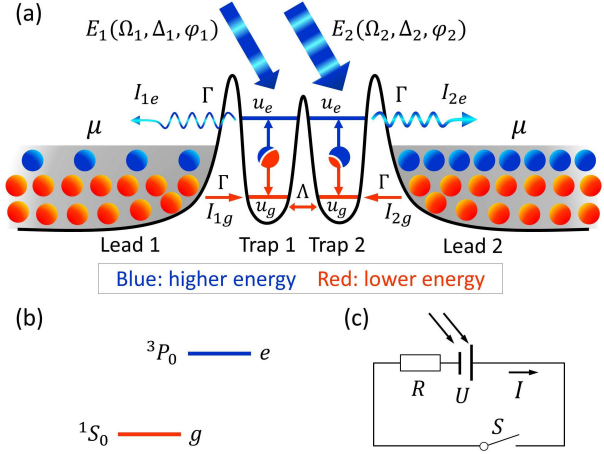}\\
\caption{(Color on line) (a) Schematic illustration of the asymmetric field photovoltaic cell. Atoms in the left and right leads have the same chemical potentials $\mu$. $E_{1}$ and $E_{2}$ represent two lights with different amplitudes, different frequencies or different phases, driving atoms in trap $1$ and trap $2$, respectively. Blue color indicates the area of higher energy and red color denotes the area of lower energy. (b) The ground and excited state levels of alkaline-earth(-like) atoms. (c) Conceptual structure of closed circuit which consists of the neutral atom photovoltaic cell with its effective voltage $U$, a resistance $R$ and a switch $S$.}
\label{sys}
\end{figure}

The model that we consider here is a double-trap opened system which is coupled to the left and right leads via atom tunneling as conceptually illustrated in Fig.~\ref{sys}(a). Atoms from the leads occupy the ground state of two traps at the same probability due to potential energy difference. Inhomogeneous external lights excite atoms in the traps into their excited states and rearrange the atom occupations. It is leads to polarization of the double-trap system. The ground and a Metastable states of Fermion alkaline-earth(-like) atoms with long lifetime clock transitions can be used in our model (See Fig.~\ref{sys}(b)). At least two traps are necessary to built the cell, since polarization of atom population as an effective bias voltage of the battery have to be constructed in the two traps.

\begin{center}
\textbf{A. Hamiltonian}
\end{center}

Optical potentials for the two traps and atomic reservoir are insensitive to the internal state of atoms~\cite{Arora}. For neutral atoms, the main inter atom interaction comes from their collisions. Each trap is set to be narrow enough and either one or no atom occupies the trap in the collisional blockade regime of strong repulsion~\cite{Recati}. In this blockade regime, the double-trap system could be described by the single-particle Hamiltonian~\cite{Gerbier,Wall}
\begin{eqnarray}
H_{trap}=\sum_{\alpha,s}\varepsilon_{s}a_{\alpha s}^{\dag}a_{\alpha s}.
\label{eq:trap}
\end{eqnarray}
Here, $\varepsilon_{s}$ represent energy of a single atom which is the same in trap $\alpha=1$ and trap $\alpha=2$, $s=g, e$ denote the ground and excited states an individual atom. $a_{\alpha s}$ ($a_{\alpha s}^{\dag}$) would annihilate (create) an atom in state $s$.

In coherent atom-light interaction, pseudospin degree of freedom in a single atom can be coupled to its momentum, it is call spin-orbit coupling\cite{XJLiu,Campbell,Jimenez}. Therefore, when a two-level atom is driven from the ground state $g$ to its excited state $e$, correspondingly its momentum change from $k$ to $k+\Delta k$, namely, $|g,k\rangle \rightarrow |e,k+\Delta k\rangle$~\cite{Gerbier,Wall,Livi,Celi,Mancini}. The momentum increase $\Delta k$ is related to synthetic magnetic
flux per plaquette $\phi$ and the potential scale $\lambda_{M}$ (magic wavelength) as $\Delta k=2\phi/\lambda_{M}$~\cite{wxlai,Wall,Livi}. In photovoltaic effect, energy transfer is more important than momentum transfer. From the point of view of energy transformation, in this process, excitation of the internal state of an atom is accompanied by its kinetic energy change. It reveals that energy of the two-level atom should be written in the form, $\varepsilon_{g}=\epsilon_{g}+u_{g}$ and $\varepsilon_{e}=\epsilon_{e}+u_{e}$, where $\epsilon_{g}$ and $\epsilon_{e}$ denote internal (electronic) level of the atom, $u_{g}=k^{2}/2m$ and $u_{e}=(k+\Delta k)^{2}/2m$ represent kinetic energy of the atom for mode $1$ and mode $2$ in a trap, respectively, where $m$ is the atomic mass (Planck's constant is taken to be $\hbar=1$ throughout this work). In this way, light energy would be transformed and stored into atomic gas in the form of internal electronic levels and the energy of atom modes.

Inter-trap coupling Hamiltonian is given by the following expression~\cite{Livi}
\begin{eqnarray}
H_{itrap}=\Lambda\sum_{s}(a_{1 s}^{\dag}a_{2 s}+H.c.).
\label{eq:trap}
\end{eqnarray}
Atoms coherently transfer to the neighboring trap through the central barrier with the rate $\Lambda$ for any states.

As reservoirs, left lead is connected to the left trap and right lead is connected to the right trap. The two leads consist of non-interacting Fermion atomic gas bounded in large optical wells with the same chemical potentials $\mu$. It means there is no chemical potential difference between the two leads. Hamiltonian of the reservoirs can be described by the energy of free atomic gas,
\begin{eqnarray}
H_{lead}=\sum_{\alpha,s,p}\varepsilon_{sp}b_{\alpha s p}^{\dag}b_{\alpha s p}.
\label{eq:lead}
\end{eqnarray}
Analogous to the energy configuration of atoms in the two traps, energy of the atomic gas in the two leads could be described in the two parts $\varepsilon_{sp}=\epsilon_{s}+u_{p}$ with $s=g,e$ for internal states. What difference is atom modes in the leads characterized by the energy $u_{p}=p^{2}/2m$ with continuously spectrum of momentum $p$. Operator $b_{\alpha s p}^{\dag}b_{\alpha s p}$ represents occupation number in lead $\alpha=1$ (or $2$) with annihilation operator $b_{\alpha s k}$ and creation operator $b_{\alpha s p}^{\dag}$. Due to the extension of atom wave function beyond the potential barriers, exchange of atom number occurs between the double-trap system and its reservoirs. Hamiltonian of the exchange interaction could be written as,
\begin{eqnarray}
H_{exch}=\sum_{\alpha,s,p}V_{p}(b_{\alpha s p}^{\dag}a_{\alpha s}+H.c.),
\label{eq:exch}
\end{eqnarray}
where, the tunneling amplitude $V_{p}$ is considered to be insensitive to the atom states $s$ and the same for the left and right leads.

Until now the whole configuration mentioned above is symmetrically designed for the perpendicular line through the system's midpoint. Asymmetry comes from the optical beams which are driving atoms in the the double-trap. In detail, two optical beams, $E_{1}$ and $E_{2}$ are applied, where $E_{1}$ is acting on the trap $1$ and $E_{2}$ is acting on the trap $2$ as shown in Fig.~\ref{sys}. The optical beams are expected to import energy into atoms, therefore, they should be running waves in the forms of $E_{1}=E_{1 0}cos(\omega_{1}t+r_{1}\cdot k_{1}+\phi_{10})$ and $E_{2}=E_{2 0}cos(\omega_{2}t+r_{2}\cdot k_{2}+\phi_{20})$, respectively. In the electric dipole approximation, two-level atoms interacting with the monochromatic fields can be written as~\cite{Scully,Meystre}
\begin{eqnarray}
H_{drive}=\sum_{\alpha}\frac{\Omega_{\alpha}}{2}(a_{\alpha e}^{\dag}a_{\alpha g}e^{-i(\omega_{\alpha}t+\varphi_{\alpha})}+H.c.),
\label{eq:drive}
\end{eqnarray}
where the Rabi frequency $\Omega_{\alpha}=\frac{|\wp_{ge}|E_{\alpha 0}}{\hbar}$ is proportional to the beam amplitude $E_{\alpha 0}$ and the absolute value of dipole matrix element $\wp_{ge}=|\wp_{ge}|e^{i\Theta}$ ($i$ is imaginary part of complex number). The phase $\varphi_{\alpha}$ in the Hamiltonian should include the beam phase $r_{\alpha}\cdot k_{\alpha}+\phi_{\alpha0}$ and the phase $\Theta$ of the dipole matrix element.

\begin{center}
\textbf{B. Equation of motion}
\end{center}

Total density matrix $\rho_{tot}$ of the whole configuration satisfies the quantum Liouville's equation $\partial\rho_{tot}/\partial t=-i[H,\rho_{tot}]$, where $H=H_{trap}+H_{itrap}+H_{drive}+H_{lead}+H_{exch}$. Using the unitary operator $e^{iH_{0}t}$, the equation of motion can be written into interaction picture,
\begin{eqnarray}
\frac{\partial\tilde{\rho}_{tot}}{\partial t}=-i[\tilde{H},\tilde{\rho}_{tot}],
\label{eq:int-equ}
\end{eqnarray}
where the free evolution Hamiltonian consists of two parts $H_{0}=H_{trap}+H_{lead}$, the density matrix in interaction picture is $\tilde{\rho}_{tot}=e^{iH_{0}t}\rho_{tot}e^{-iH_{0}t}$ and the Hamiltonian in interaction picture is $\tilde{H}=\tilde{H}_{itrap}+\tilde{H}_{drive}+\tilde{H}_{exch}$. In detail, each term can be written as
\begin{eqnarray}
\tilde{H}_{itrap}&=&\Lambda\sum_{s}a_{1 s}^{\dag}a_{2 s}+H.c.,
\label{eq:intitrap}
\end{eqnarray}
\begin{eqnarray}
\tilde{H}_{drive}&=&\sum_{\alpha}\frac{\Omega_{\alpha}}{2}a_{\alpha e}^{\dag}a_{\alpha g}e^{i(\Delta_{\alpha}t-\varphi_{\alpha})}+H.c.,
\label{eq:intdrive}
\end{eqnarray}
\begin{eqnarray}
\tilde{H}_{exch}&=&\sum_{\alpha,s,p}V_{p}b_{\alpha s p}^{\dag}a_{\alpha s}e^{i(\varepsilon_{sp}-u_{s})t}+H.c.,
\label{eq:intexch}
\end{eqnarray}
where $\Delta_{\alpha}=\varepsilon_{e}-\varepsilon_{g}-\omega_{\alpha}$ denotes the atom-light detunings in trap $\alpha=1$ and trap $\alpha=2$. One can substitute the integration of Eq. \eqref{eq:int-equ} into itself and reach

\begin{eqnarray}
\frac{\partial\tilde{\rho}_{tot}(t)}{\partial t}&=&-i[\tilde{H}_{itrap}(t)+\tilde{H}_{drive}(t),\tilde{\rho}_{tot}(t)] \notag\\
&&-i[\tilde{H}_{exch}(t),\tilde{\rho}_{tot}(0)] \notag\\
&&-\int_{0}^{t}[\tilde{H}_{exch}(t),[\tilde{H}(t'),\tilde{\rho}_{tot}(t')]]dt'.
\label{eq:int-equ2}
\end{eqnarray}

Atomic gas in lead $1$ and lead $2$ can be seen as large reservoirs of atoms in equilibrium state with a great number of microstates. Actually, the total density matrix could be written as $\rho_{tot}(t)=\rho(t)\rho_{1}\rho_{2}$, where $\rho$ is density matrix of the double-trap system, $\rho_{1}$ and $\rho_{2}$ are time independent density matrices of the two leads, respectively. As a density matrix of the sub-system, $\rho$ satisfies the following equation in interaction picture,
\begin{eqnarray}
\frac{\partial\tilde{\rho}(t)}{\partial t}&=&-i[\tilde{H}_{itrap}(t)+\tilde{H}_{drive}(t),\tilde{\rho}(t)] \notag\\
&&-\int_{0}^{t}Tr[\tilde{H}_{ex}(t),[\tilde{H}_{ex}(t'),\tilde{\rho}(t')\rho_{1}\rho_{2}]] dt',
\label{eq:int-equ3}
\end{eqnarray}
where $\tilde{\rho}(t)=Tr[\tilde{\rho}_{tot}(t)]$ and Tr represents taking trace over all microstates of the two leads. For the density matrix $\rho_{\alpha}$ of uncorrelated thermal equilibrium atomic gas in the leads, the terms $Tr[b_{\alpha s p}^{\dag}b_{\alpha s p}\rho_{\alpha}]=f_{\alpha}(\varepsilon_{sp})$ in Eq. \eqref{eq:int-equ3} is actually the mean occupation number of a single particle state, namely, the Fermi-Dirac distribution function, $f_{\alpha}(\varepsilon_{sp})=\frac{1}{e^{(\varepsilon_{sp}-\mu)/k_{B}T}+1}$, where $k_{B}$ is the Boltzmann constant and $T$ is the temperature of the leads.

In the end, using Born-Markov approximation for the time integration in Eq. \eqref{eq:int-equ3} and transforming the equation back to Schr\"{o}inger picture with the unitary operator $e^{-it\Sigma_{\alpha}\Delta_{\alpha}a_{\alpha e}^{\dag}a_{\alpha e}}$, we can obtain equation of motion of the system~\cite{Meystre,Scully},
\begin{eqnarray}
\frac{\partial\rho}{\partial t}=-i[H_{cell},\rho]+\sum_{\alpha=1,2}\mathcal{L}_{\alpha}\rho.
\label{eq:equation}
\end{eqnarray}
The first term on the right side is free evolution of the double-trap system, the photovoltaic cell component, with the effective Hamiltonian $H_{cell}=\Sigma_{\alpha}\Delta_{\alpha}a_{\alpha e}^{\dag}a_{\alpha e}+\sum_{s}\Lambda(a_{1s}^{\dag}a_{2s}+H.c.)+\sum_{\alpha}\frac{\Omega_{\alpha}}{2}(a_{\alpha g}^{\dag}a_{\alpha e}e^{i\varphi_{\alpha}}+H.c.)$. The second term indicates the exchange of atoms between the double-trap system and the leads. The Lindblad super operators acting on the density matrix $\rho$ can be written as $\mathcal{L}_{\alpha}\rho=\frac{\Gamma}{2}\sum_{s}[f_{\alpha}(u_{s})(a_{\alpha s}^{\dag}\rho a_{\alpha s}-a_{\alpha s}a_{\alpha s}^{\dag}\rho)+(1-f_{\alpha}(u_{s}))(a_{\alpha s}\rho a_{\alpha s}^{\dag}-a_{\alpha s}^{\dag}a_{\alpha s}\rho)+H.c.]$ at finite temperature $T$. Detail expression of the atom exchange rate is written in the state independent form $\Gamma=2\pi|V_{p}|^{2}D(\varepsilon_{sp})$ under adiabatic approximation ($D(\varepsilon_{sp})$ is density of states of atom).

Each trap has three basic states, occupation of the ground state atom $|\alpha, g\rangle$, occupation of the excited state atom $|\alpha, e\rangle$ and empty state $|\alpha, 0\rangle$. Considering the two traps, they have all together nine basic states $\{|1,s_{1}\rangle|2,s_{2}\rangle, s_{1},s_{2}=0, g, e\}$. In the Hilbert space of these $9$ basic states, equation of motion for the atomic density matrix elements can be calculated. Using the density matrix of system, probabilities of the states of each trap can be achieved $P_{10}=tr\{a_{1g}a_{1g}^{\dag}\rho\}$, $P_{20}=tr\{a_{2g}a_{2g}^{\dag}\rho\}$, $P_{1g}=tr\{a_{1g}^{\dag}a_{1g}\rho\}$, $P_{2g}=tr\{a_{2g}^{\dag}a_{2g}\rho\}$, $P_{1e}=tr\{a_{1e}^{\dag}a_{1e}\rho\}$ and $P_{2e}=tr\{a_{2e}^{\dag}a_{2e}\rho\}$, which describe probabilities of the empty trap, the ground state atom occupation and the excited atom occupation in trap $1$ and trap $2$, respectively. The tr here represents trace over all states of the double-trap system.

Polarization of atom occupation in the double-trap leads to atomic current in the system. Atomic current here is defined that the number of atoms passing a cross section of trap-lead contact within unit time. The charge of a single neutral atom could be defined as $q=1$, then, the unit of atomic current is $s^{-1}$, where $s$ represent one second. The current of atoms may be calculated using the continuity equation\cite{Davies,Jauho,Twamley}:
\begin{eqnarray}
q\frac{d N(t)}{dt}=I_{1}-I_{2},
\label{eq:current}
\end{eqnarray}
where, $N(t)=\Sigma_{\alpha,s}tr\{a_{\alpha s}^{\dag}a_{\alpha s}\rho(t)\}$ is the occupation number of atoms in the double-trap system at time $t$. Here, direction of the positive current is defined to be from left to right. On the right side of the equation, $I_{1}$ represents the left current in lead $1$, $I_{2}$ denotes the right current in lead $2$. Actually, the left and right current consist of the ground state current $I_{\alpha e}$ and excited state current $I_{\alpha e}$, namely $I_{1}=I_{1g}+I_{1e}$ and $I_{2}=I_{2g}+I_{2e}$. Furthermore, substituting Eq. \eqref{eq:equation} into the continuity equation\eqref{eq:current}, expressions of ground state atomic current and excited state atomic current at lead $1$ and lead $2$ can be achieved separately, which gives rise to
\begin{eqnarray}
I_{1g}=q\Gamma [f_{1}(u_{g})P_{10}-(1-f_{1}(u_{g}))P_{1g}],
\label{eq:ige1}
\end{eqnarray}
\begin{eqnarray}
I_{1e}=q\Gamma [f_{1}(u_{e})P_{10}-(1-f_{1}(u_{e}))P_{1e}],
\label{eq:ige2}
\end{eqnarray}
\begin{eqnarray}
I_{2g}=-q\Gamma [f_{2}(u_{g})P_{20}-(1-f_{2}(u_{g}))P_{2g}],
\label{eq:ige3}
\end{eqnarray}
and
\begin{eqnarray}
I_{2e}=-q\Gamma [f_{2}(u_{e})P_{20}-(1-f_{2}(u_{e}))P_{2e}].
\label{eq:ige4}
\end{eqnarray}
The continuity equation\eqref{eq:current} reveals $I_{1}=I_{2}$ in stationary state circuit. Then, we have the averaged total current $I=(I_{1}+I_{2})/2$ in the form,
\begin{eqnarray}
I=\Gamma \sum_{\alpha,s}(-1)^{\alpha-1}[f_{\alpha}(u_{s})P_{\alpha0}-(1-f_{\alpha}(u_{s}))P_{\alpha s}]
\label{eq:total-current}
\end{eqnarray}
which represent the net current through the cell. Here, $q=1$ is considered for a neutral atom.

\begin{center}
\textbf{III. RESULTS}
\end{center}

When energy scales satisfy the relation $\mu-u_{g}, u_{e}-\mu \gg k_{B}T$, the Fermi-Dirac distribution functions tend to their extreme values $f_{1}(u_{g})=f_{2}(u_{g})\approx 1$, $f_{1}(u_{e})=f_{2}(u_{e})\approx 0$. Based on the low temperature limit and the representation of atom state probabilities, one can have the total net current $I$ as follows
\begin{eqnarray}
I=\frac{\Gamma}{2}[(P_{2e}-P_{1e})+(P_{10}-P_{20})],
\label{eq:neti}
\end{eqnarray}
It would be demonstrated numerically in the following that the probabilities of empty traps always satisfy the conditions $P_{10}\approx P_{20}$ and $|P_{10}-P_{20}|\ll |P_{2e}-P_{1e}|$, then Eq.\eqref{eq:neti} can be written in a simple form,
\begin{eqnarray}
I=\frac{\Gamma}{2}(P_{2e}-P_{1e}).
\label{eq:simpneti}
\end{eqnarray}
In this equation, $P_{2e}-P_{1e}$ represent the effective charge collection $Q$ in the cell and $2/\Gamma$ is characteristic time $\tau$ for atom transfer through the double-trap system. Therefore, the current can be written as $I=Q/\tau$ similar to the concept of electronic current.

Next, let us pay attention to more detail properties of the photovoltaic cell. In Eq.\eqref{eq:equation}, amplitudes $E_{\alpha0}$, frequencies $\omega_{\alpha}$, phases $\varphi_{\alpha}$ of the two applied fields ($\alpha=1,2$) are reflected in the system Hamiltonian $H_{cell}$ in the forms of Rabi frequencies $\Omega_{\alpha}$, atom-light detunings $\Delta_{\alpha}$, and phases $\varphi_{\alpha}$, respectively. Next, affects on the atoms from these parameter differences would be discussed in the following. In the Hilbert space of the nine basic states mentioned above, density matrix of the system can be achieved solving Eq. \eqref{eq:equation} in stationary condition $\partial\rho/\partial t=0$, considering the relation $Tr\{\rho\}=1$ of normalization. The basic parameters are $\Gamma=2\pi\times400Hz$, $\Lambda=2\pi\times800Hz$, $k_{B}T=\Gamma/10$, $\varepsilon_{g}=0$ and $\mu-u_{g}, u_{e}-\mu \gg k_{B}T$. Based on these conditions, numerical results are given in the following sections.

\begin{figure}
  \includegraphics[width=8.5cm]{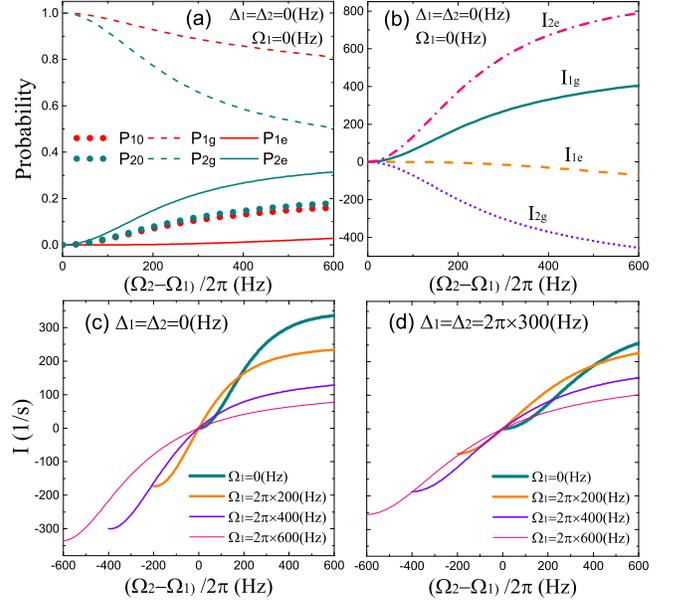}\\
  \caption{(Color on line) (a) Probabilities of trap states as a function of Rabi frequency difference $\Omega_{2}-\Omega_{1}$ at resonant coupling $\Delta_{\alpha}=0$. (b) The ground and excited state currents versus the Rabi frequency difference $\Omega_{2}-\Omega_{1}$ at resonant coupling $\Delta_{\alpha}=0$. (c) Total current as a function of Rabi frequency difference $\Omega_{2}-\Omega_{1}$ at resonant coupling $\Delta_{\alpha}=0$. (d) Total current as a function of Rabi frequency difference $\Omega_{2}-\Omega_{1}$ with detuning $\Delta_{\alpha}=2\pi\times300 Hz$. Other parameters used here are $\varphi_{1}=\varphi_{2}=0$.}\label{iomg}
\end{figure}

\subsection{Different amplitudes $E_{10}\neq E_{20}$}

The most simple case is that a single resonant light is used to trap $2$ and there is no light in trap $1$, $\Omega_{1}=0$. As illustrated in Fig.~\ref{iomg} (a), along with increase of the light strength in trap $2$, probability of excited atom $P_{2e}$ in trap $2$ would remarkably increase, which is much larger than the corresponding probability $P_{1e}$ in trap $1$. The fact induces polarization of the two traps, $P_{2e}>P_{1e}$ which determine the charge collection $Q$ in this battery. Therefore, the photovoltaic cell works under single light beams actually.

Due to trap-lead exchange tunneling and external light driving, the system gives nonzero probabilities of empty traps $P_{10}$ and $P_{20}$. Although they are appeared to be not important for determination of the current $I$ because of $P_{10}\approx P_{20}$, in fact they are very significant for keeping steady current in the system. As plotted in Fig.~\ref{iomg} (a) and (b), probabilities of empty traps $P_{10}$ and $P_{20}$ really give rise to the ground state atom current. Therefore, the states of empty traps play the role of 'hole' states in semiconductors~\cite{YYu}. The ground state atoms atoms play the role of electrons in filled valence band, excited state atoms play the role of electrons in conduction band in semiconductor quantum dot based electronic solar cell~\cite{Luque,Al-Ahmadi}.

\begin{figure}
  \includegraphics[width=8.5cm]{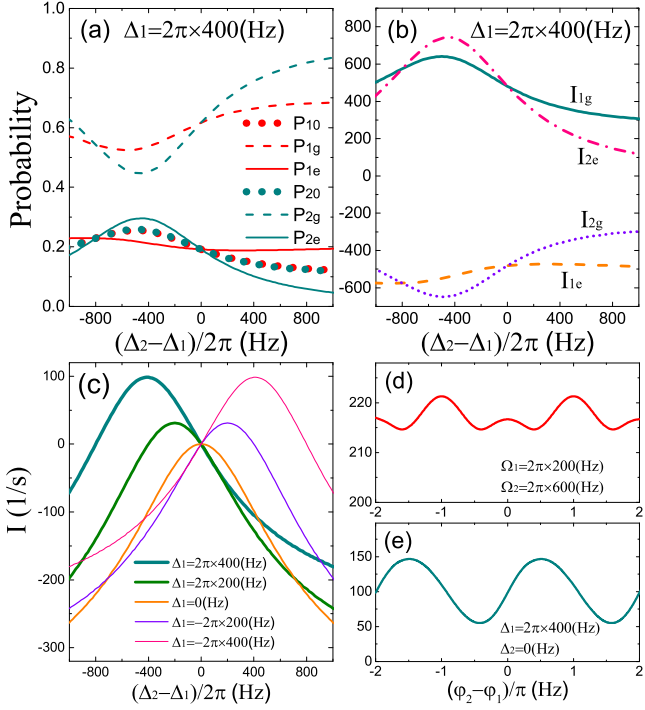}\\
  \caption{(Color on line) (a) Trap states probabilities as a function of atom-light detuning difference $\Delta_{2}-\Delta_{1}$. (b) The ground and excited state currents versus the detuning difference $\Delta_{2}-\Delta_{1}$. (c) Total current as a function of the detuning difference $\Delta_{2}-\Delta_{1}$. The common parameters used in above figures are: $\Omega_{1}=\Omega_{2}=2\pi\times600Hz$ and $\varphi_{1}=\varphi_{2}=0$. (d) Total current as a function of phase difference $\varphi_{2}-\varphi_{1}$ under different Rabi frequencies $\Omega_{1}$ and $\Omega_{2}$ with the parameter $\Delta_{1}=\Delta_{2}=0Hz$. (e) Total current change for the phase difference $\varphi_{2}-\varphi_{1}$ under different detunings $\Delta_{1}$ and $\Delta_{2}$ with $\Omega_{1}=\Omega_{2}=2\pi\times600Hz$.}\label{idelta}
\end{figure}

The more general case is both two lights are applied to the atoms, contributing nonzero amplitudes $E_{10}$ and $E_{20}$. Currents in this situation ($\Omega_{1}\neq\Omega_{2}$) are plotted in Fig.~\ref{iomg} (c) and (d) under the resonant and non-resonant couplings. When $\Omega_{1}=\Omega_{2}$, current would always be zero. It is known from Fig.~\ref{idelta} (d) that relative phase $\varphi_{2}-\varphi_{2}$ between light $E_{1}$ and $E_{2}$ does not obviously affect the current behavior. The fact reveals an important feature of this system, the cell works even under short coherent light.

\subsection{Different frequencies $\omega_{1}\neq \omega_{2}$}

When frequencies $\omega_{1}$ and $\omega_{2}$ of the two lights are different, which is reflected in the detuning difference $\Delta_{2}-\Delta_{1}$, probability $P_{1e}$ of an excited atom in trap $1$ is higher or lower than that $P_{2e}$ in trap $2$ as demonstrated in Fig.~\ref{idelta} (a). It is the double-trap polarization in atomic state distribution. The atom distribution probabilities directly determine current of atoms at different state and directions (See Fig.~\ref{idelta} (b)). double-trap polarization $P_{2e}-P_{1e}$ (or $P_{2g}-P_{1g}$) is disappeared when detunings satisfy $|\Delta_{1}|=|\Delta_{2}|$. The fact is also reflected in the total current in which current would reduced to be zero at two points as shown in Fig.~\ref{idelta} (c).

For the change of relative phase $\varphi_{2}-\varphi_{2}$ between light $E_{1}$ and $E_{2}$, current fluctuate around a positive value as illustrated in Fig.~\ref{idelta} (e). It reveals nonzero mean current can be preserved even under random phase difference of the two light.

\begin{figure}
  \includegraphics[width=8.5cm]{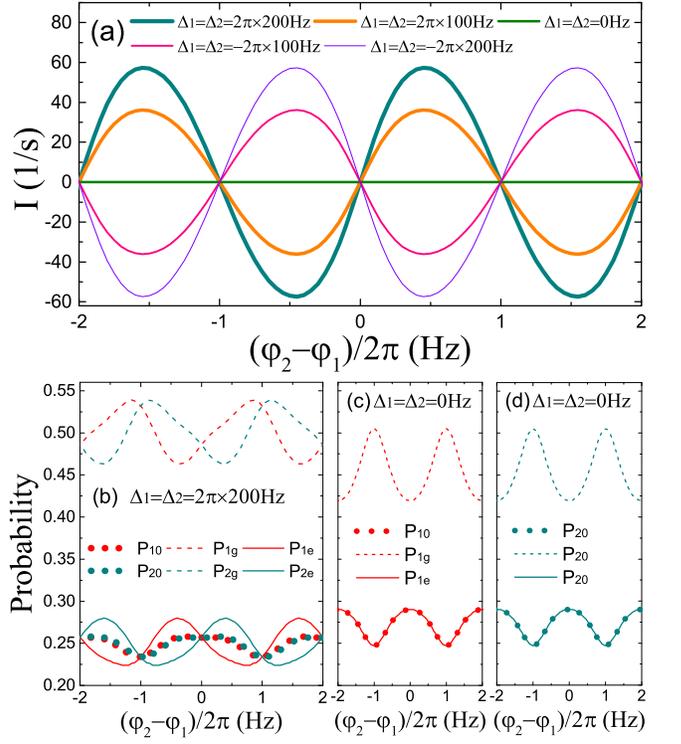}\\
  \caption{(Color on line) (a) Current as a function of phase difference $\varphi_{2}-\varphi_{1}$ under several values of detunings. (b), (c) and (d) Trap states probabilities plotted under the change of relative phase in the case of resonant and off-resonant couplings. The common parameters used here are $\Omega_{1}=\Omega_{2}=2\pi\times600Hz$.}\label{iphi}
\end{figure}

\subsection{Different phases $\varphi_{1}\neq \varphi_{2}$}

Now, the two fields $E_{1}$ and $E_{2}$ have the same amplitudes and frequencies, but they have different phases. Interference of these
two fields reflected in the double-trap system with current fluctuation as shown in Fig.~\ref{iphi} (a). The current behavior is directly related to the trap states distribution, especially the double-trap polarization od excited atom distribution $P_{2e}-P_{1e}$ (See Fig.~\ref{iphi} (b)). It is different from the case of amplitude and frequency asymmetry, current should be disappeared if the two fields $E_{1}$ and $E_{2}$ are incoherent with random phases, since the current versus their relative phase fluctuates near the zero current.

Fig.~\ref{iphi} (a) displays two interesting features. For one, current is in opposite direction for red detuning $\Delta_{\alpha}>0$ and blue detuning $\Delta_{\alpha}<0$. For the other, current disappears when lights resonantly interact with atoms $\Delta_{\alpha}=0$. It may be interpreted in the theory of wave beat. It is well known that superposition of two waves with near frequencies leads to wave beat effect. When $\Delta_{\alpha}\neq0$, atom-light coupling could create wave beat of atom waves~\cite{Decamps,Villavicencio}. It can interpret the fact that if current directions are opposite for $\Delta_{\alpha}>0$ and $\Delta_{\alpha}<0$. Indeed, when  $\Delta_{\alpha}=0$, beat would be disappeared and atom waves in the double-trap form standing wave which could not induces current and could not transfer energy (See Fig.~\ref{iphi}). Furthermore, the standing wave also does not have to do with the phases of the two lights. Therefore, net atom current require $\Delta\neq0$, which is in fact the propagation of atomic beat wave.

\begin{center}
\textbf{IV. FEASIBILITY}
\end{center}

Optical traps with both narrow and wide scales have been implemented in experiments for atom controlling and manipulation~\cite{McKeever,Thompson,Reiserer}. Bounding and controlling a few atoms in tight optical traps are also achievable using ultra thin laser beams with high Fidelity~\cite{Recati,Caliga2,Caliga3,Cooper}. First choice of neutral atoms are alkaline-earth (like) metals, such as $^{87}Sr$~\cite{Cooper}, $^{171}Yb$~\cite{Gorshkov}, $^{174}Yb$~\cite{Hara} and $^{40}Ca$~\cite{Kraft}, because they have long lived excited states due to the clock transition $^{1}S_{0}(g)-^{3}P_{0}(e)$. Since energy would be stored in the excited state of atoms, coherent lifetime of atoms is very important. Lifetime of the clock states in these alkaline-earth (like) atoms, reported to be from a few seconds to a few tens seconds~\cite{Daley}. The time scales including lifetime of atoms, instability of optical potentials and atom collision induced incoherence are limited, they can be avoided as soon as the time scale is longer than the time for operation of light gate and probes. Furthermore, it is demonstrated above that for the photovoltaic cell of amplitude difference and frequency difference, demand of the light coherence is very low.

\begin{center}
\textbf{V. CONCLUSIONS}
\end{center}

In conclusions, we show that two different light beams driving atoms in a double-trap opened system create charging of neutral atoms, converting light energy into the atomic energy through both resonant and off-resonant atom-light couplings. The photovoltaic cell mainly have four characteristics. Firstly, energy transferred from external light would be stored both in internal states of atoms and atom modes states. Secondly, effective charge collection at unit time represent current of atoms. Thirdly, currents due to amplitude differences and frequency differences are several times to ten times larger than that created by the phase differences of the external fields. Finally, The cell can works any long time in principle, since the steady current in the system is insensitive to phase change of external lights. Although this mechanism is demonstrated based on alkaline-earth (like) atoms, it should applied to other particles charged or uncharged. Our model may have potential applications on energy transformation and storage, especially on the development of neutral particle circuit.

\begin{acknowledgments}
This work was supported by the Scientific Research Project of Beijing Municipal Education Commission (BMEC) under Grant No.KM202011232017, supported by the Research Foundation of Beijing Information Science and Technology University under Grant No. 1925029, also supported by the National Key R and D Program of China under grants No.2016YFA0301500, NSFC under grants No.61835013, Strategic Priority Research Program of the Chinese Academy of Sciences under grants Nos.XDB01020300, XDB21030300.
\end{acknowledgments}

\end{document}